\documentclass[arguments,manuscript]{aastex631}
\usepackage{subfigure}
\shorttitle{The power of velocity oscillations on the magnetic field inclination}
\shortauthors{Koyama ans Shimizu}
\graphicspath{{./}{figures/}}

\begin{document}

\title{Solar chromospheric heating by magnetohydrodynamic waves: dependence on magnetic field inclination}

\author{Mayu Koyama}
\affiliation{Department of Earth and Planetary Science, The University of Tokyo, \\
7-3-1 Hongo, Bunkyo-ku, Tokyo 113-0033, Japan}
\affiliation{Institute of Space and Astronautical Science, Japan Aerospace Exploration Agency, \\
3-1-1 Chuo-ku, Sagamihara, Kanagawa 252-5210, Japan}

\author[0000-0003-4764-6856]{Toshifumi Shimizu}
\affiliation{Institute of Space and Astronautical Science, Japan Aerospace Exploration Agency, \\
3-1-1 Chuo-ku, Sagamihara, Kanagawa 252-5210, Japan}
\affiliation{Department of Earth and Planetary Science, The University of Tokyo, \\
7-3-1 Hongo, Bunkyo-ku, Tokyo 113-0033, Japan}








\begin{abstract}

A proposed mechanism for solar chromospheric heating is 
that magnetohydrodynamic waves 
propagate upward along 
magnetic field lines and 
dissipate their energy in the chromosphere. 
In particular, compressible magneto-acoustic waves 
may contribute to the heating. 
Theoretically, the components 
below the cutoff frequency cannot propagate into the chromosphere; 
however, the cutoff frequency depends on the inclination 
of the magnetic field lines. 
In this study, using high temporal cadence spectral data of IRIS and Hinode SOT 
spectropolarimeter (SP) in plages, 
we investigated the dependence of the low-frequency waves 
on 
magnetic-field properties
and quantitatively estimated the amount of energy dissipation
in the chromosphere.
The following results were obtained: (a) The amount of energy dissipated 
by the low-frequency component (3--6 mHz) increases with the field inclination 
below 40 degrees, whereas it is decreased as a function of
the field inclination above 40 degrees.
(b) The amount of the energy is enhanced toward $10^4\ \rm W/m^2$,  which is the energy required for 
heating in the chromospheric plage regions, when the magnetic field is higher than 600 G and inclined more than 40 degree. 
(c) In the photosphere, the low-frequency component has much more power in the magnetic field inclined more and
weaker than 400 G.
The results suggest that 
the observed low-frequency components 
can bring
the energy along the magnetic field lines and that
only a specific range of the field inclination angles and field strength 
may allow the low-frequency component to 
bring the sufficient amount of the energy into the chromosphere.

\end{abstract}



\section{Introduction} \label{sec:intro}

The solar chromosphere is the dynamical and thin atmosphere formed 
above the solar surface (photosphere); it has a temperature of approximately
10,000 K, higher than that of the photosphere (6,000 K). 
The amount of energy required to retain the chromospheric temperature
is $10^4\ \rm W/m^2$ for active regions, an order of magnitude larger than 
the heating for the solar corona, 
which is the lower-density atmosphere heated to over 1 MK beyond
the magnetic structures of chromosphere \citep{Withbroe1977}.
The physical mechanisms
responsible for transferring the energy to the chromosphere and the corona 
and dissipating it there are still in debate.
The major candidates are magnetohydrodynamic (MHD) waves and
nanoflares. 
The turbulent convective gas motions at the photosphere excite
MHD waves propagating to the chromosphere and the corona
along the magnetic field lines \cite[e.g., a review by][and references therein]{Jess2023}, 
whereas they generate tangential discontinuity and 
braiding in magnetic field lines in the corona, leading to 
numerous tiny magnetic reconnection events
\cite[e.g., a review by][and references therein]{Pontin2020}, 
so-called nanoflares \citep{Parker1988}. 
Both mechanisms are closely associated with dynamical behaviors of the magnetic field lines
in the solar atmosphere.

Compressible magneto-acoustic waves may be evolved to shock waves due to steepening, and their dissipation may contribute to the heating of the atmosphere. 
Therefore, investigating the propagation of magneto-acoustic waves in the low atmosphere
to understand their roles in chromospheric heating is important. There are two types of magneto-acoustic waves: fast mode---in which the phase relation between the gas pressure and magnetic pressure is in-phase---and slow mode---in which the phase relation between the gas pressure and magnetic pressure are in the opposite phase. The slow mode propagates along magnetic field lines. 
In the presence of a magnetic field, sunspots are the most apparent regions for typical display of 
oscillations in intensity and Doppler 
velocity. The power spectra are dominated for 5 min (approximately 3 mHz) in the photosphere and
3 min (approximately 5 mHz or higher) in the chromosphere \citep{Centeno2009}.
The oscillations in the umbrae and penumbrae of sunspots have been generally interpreted as MHD waves.
\cite{Kanoh2016} conducted simultaneous observations of 
the Hinode and IRIS satellites and showed the existence of the slow mode waves in a sunspot umbra based on 
the phase relation of waves observed at the photosphere and the chromosphere. 
In addition, the energy flux in each layer was derived and the energy dissipation was estimated from the difference. They obtained an energy flux of $8.3\times10\ \rm W/m^2$ for waves in the upper chromosphere and $2.0\times10^4\ \rm W/m^2$ for the photosphere, demonstrating that the difference is sufficient to heat the chromosphere. 
Oscillating signatures in sunspot penumbrae, known as running penumbral waves, are interpreted
as upward propagating slow-model waves guided by the magnetic field lines \citep{Lohner-Bottcher2015}. 
The dominant frequency of the waves changes toward lower values as 
one moves from the umbral center to the surroundings where the magnetic fields
are more inclined to the surface \citep{Bloomfield2007, Jess2013}. 

The critical frequency at which a wave can propagate is called the cutoff frequency; it is defined 
as $\nu_{ac}=\frac{\gamma g}{4\pi c_s}$, 
where $c_s$ is the sound speed, $\gamma$ is the specific heat ratio of a monatomic molecule, and $g$ is the gravitational acceleration (\citealt{Priest2014}). 
The cutoff frequency is at its maximum in the lowest temperature layer, i.e., the temperature minimum,
clearly shown as the height variation of the cutoff frequency in observations of a sunspot umbra \citep{Felipe2018}
and in numerical models \citep{Felipe2020}. 
The low-frequency waves below the frequency 
cannot propagate into the chromosphere. 
In the quiet Sun, where the plasma $\beta$ is much larger than 1, the cutoff frequency is approximately 5.2 mHz ($\gamma=5/3,\ g=274\rm m/s,\it c_s=\rm7\ km/s$). \cite{Bel1977} showed theoretically that the cutoff frequency depends on the structure of the magnetic field. In the region of strong magnetic field ($\beta<1$), the cutoff frequency depends on the inclination angle of the magnetic field lines with respect to the solar surface. This is because the effective gravity on the magnetic field lines changes, and the cutoff frequency becomes
\begin{equation}
\label{Eqcutoff}
\nu_{ac}=\frac{\gamma g\cos\theta}{4\pi c_s}.
\end{equation}
Note that the cutoff frequency given in equation (\ref{Eqcutoff}) is only applicable in an isothermal atmosphere
and different forms of the cut-off frequency are inferred from various representations of
the wave equation depending on choice of variables in more general atmospheres \citep{Schmitz1998}.
\cite{McIntosh2006} observed that low-frequency magneto-acoustic waves below the cutoff frequency (5.2 mHz) propagate into the chromosphere in the sunspot penumbra 
where the magnetic field is highly tilted to the surface.
Numerical simulations of wave propagation in the low atmosphere also show that 
the magnetic field inclination is crucial for the propagation of low-frequency waves 
through dynamic magnetic structures \citep{Heggland2011} 
and wave energy flux \citep{Schunker2006, Cally2013}.

For observationally investigating the significance of wave contributions to the upper atmosphere, 
quantitatively estimating the amount of energy magneto-acoustic waves
transported from the photosphere to the chromosphere and deposited in the chromosphere
considering the magnetic field environment is essential. However, the number of studies on 
magnetic plages outside sunspots is largely limited.
\cite{Sobotka2016} used the Interferometric Bidimensional Spectrometer 
\cite[IBIS; ][]{Cavallini2006}
at the Dunn Solar Telescope (DST) to evaluate the deposited acoustic energy flux from
the power spectra of Doppler oscillations measured in the $\rm Ca\ I\hspace{-.1em}I$ 853.2 nm line core
in comparison to the radiative loss. 
In active areas around sunspots, the amount of energy dissipated by 4--9 mHz waves was estimated
to be up to $\rm5\times10^3\ W/m^2$ in the height range 1000--1500 km from the solar 
surface (the middle chromosphere).
The estimated flux is slightly insufficient to heat the chromosphere; however,
the heating contribution was revealed to increase from 23 \%\ in chromosphere 
network to 54 \%\ in a plage.
\cite{Abbasvand2020b} used observations of the $\rm Ca\ I\hspace{-.1em}I$ 854.2 nm, H$\alpha$ 
and H$\beta$ lines with the Fast Imaging Solar Spectrograph 
\cite[FISS; ][]{Chae2013} at the 1.6-m Goode Solar Telescope and the echelle spectrograph
attached to the German Vacuum Tower Telescope \citep{vonderLuhe1998}
and derived that the acoustic energy flux at frequencies up to 20 mHz 
deposited in the heights from 1000 to 1400 km can be balanced by the radiative loss
in a quiet region, but the deposited acoustic flux is insufficient in the upper chromosphere higher
than 1400 km in both quiet and plage regions. 
\cite{Abbasvand2020a} used observations of the $\rm Ca\ I\hspace{-.1em}I$ 854.2-nm line
with the IBIS and derived the acoustic energy flux in the plage, demonstrating
that it contributes by $50\%\--60$\%\ in locations with vertical magnetic field and $70\%\--90$\%\
in regions where the magnetic field is inclined more than $50^\circ$ to the solar surface normal.

In this study, we investigate roles of magnetic field properties in oscillations
observed in plages, which
are observed brightly in the chromosphere; consequently, heating is inferred to be relatively high
so that we could understand more details of energy transport and dissipation by waves.
For quantitative understanding, we estimate the energy dissipation 
in the chromosphere by comparing the energy flux at three altitudes in the plages: 
the photosphere, lower chromosphere, and upper chromosphere. 
In addition, as the cutoff frequency varies depending on the magnetic-field structure of the photosphere, 
we estimate the energy dissipation in the chromosphere 
including the energy flux due to low-frequency waves (less than 5 mHz) below the cutoff frequency. 
Determining the inclination of the magnetic field lines with respect to the solar surface with the Hinode SOT/SP, we verify whether low-frequency waves below the cutoff frequency propagate into the chromosphere depending on the inclination of the magnetic field lines. We describe observational methodologies in Section 2. 
Section 3 shows the observational results, and Section 4 discusses the interpretation of the observation results. 
A summary of this paper is given in Section 5.

\section{Observations and data analysis} \label{sec:Obs}

\begin{figure}[ht]
 \begin{center}
 \includegraphics[keepaspectratio, scale=0.7]{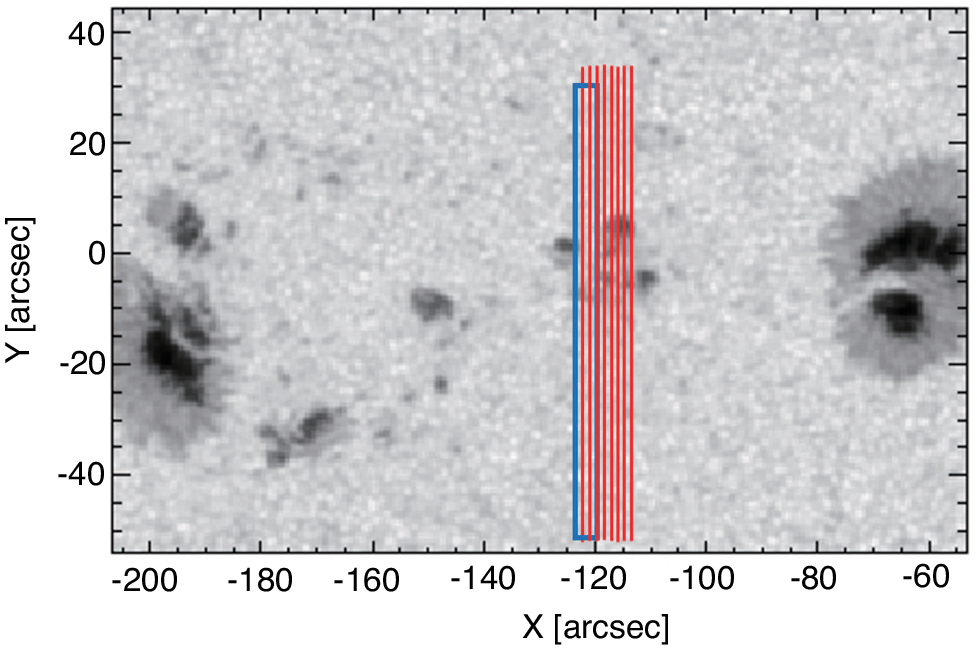}
 \end{center}
 \caption{Observation area on 
 10 February 2018.
 Blue frame shows the field of view 
 captured by the Hinode SOT observation, whereas 
 red lines show the slit positions for the IRIS observation. 
 The IRIS slit positions are 8 steps at $1''$ intervals.
 The SOT field of view contains 10 slit positions of the spectro-polarimeter measurement with $0.3"$ interval.
 }\label{FOV}
\end{figure}

Simultaneous observations of the plage region were performed by the Solar Optical Telescope (SOT) 
\citep{Tsuneta2008, Suematsu2008, Shimizu2008, Ichimoto2008} 
onboard the Hinode satellite (\citealt{Kosugi2007}) and 
the IRIS satellite (\citealt{DePontieu2014}) for approximately one hour each day on February 9--12, 2018. 
The observations were carried out along with the observation proposal named 
IRIS-Hinode Operation Plan (IHOP) 0341.
The period of each simultaneous Hinode and IRIS observation and the heliocentric coordinate of the center of the leftmost IRIS slit are shown in Table \ref{timepos}. 
Each observation region covers a plage region between the leading and following sunspots 
in active region NOAA 12699.
The observation area on February 10, 2018, as an example, is shown in Figure \ref{FOV}. 
The blue frame shows the SOT field of view, whereas the red lines show the slit positions of IRIS.

\begin{table}[ht]
 \caption{Coordinated observation period and coordinate}
 \label{timepos}
 \centering
 \begin{tabular}{c|c|c}
 \hline
 Date & Time (UT) & Position $(x,y)$  \\
  \hline\hline
 09 February 2018 & 12:59:10 -- 14:09:58 &  $(-335'',-17'')$     \\
 10 February 2018 & 11:35:13 -- 12:44:33 &  $(-122'',-9'')$        \\
 11 February 2018 & 12:05:11 -- 13:09:32 &  $(104'',-4'')$         \\ 
 12 February 2018 & 01:05:11 -- 02:14:32 &  $(226'',-5'')$         \\
  \hline 
  \end{tabular}
\end{table}

\subsection{Hinode SOT observation and analysis}

The SOT observations were carried out by a spectro-polarimeter (SP)
 (\citealt{Lites2013}) that recorded four Stokes $\rm(I,Q,U,V)$ profiles of 
 $\rm Fe\ I$ lines at $\rm 6301.5\AA$ and $\rm 6302.5\AA$ 
 with a spectral sampling of $\rm21.55\ m\AA$. 
 The field of view is $3''\times82''$ with 10 slit positions for a $3''$ width 
 at $0.3''$ interval. 
The Stokes parameters were measured with exposures during the continuous rotation of 
the polarization modulator in 1.6 s at each slit position, 
and it takes approximately 21 s to scan the file of view. 
This is the fast-mapping mode, wherein the exposures were accumulated 
at two slit positions with a slit width of $0.15''$ ; the two pixels were summed 
in the slit direction to provide spectral data with a pixel size of $0.32''$. 
In this analysis, we used the level-2 data from 
the Community Spectro-polarimetric Analysis Center (CSAC) at HAO/NCAR
\footnote{https://www2.hao.ucar.edu/csac}. 
The data was calibrated with the SP\_PREP routine in Solar SoftWare (SSW) \citep{Lites2013b}, followed
by an inversion assuming a Milne--Eddington atmosphere with 
the HAO MERLIN code to obtain physical quantities such as magnetic field strength, 
azimuth angle, and tilt relative to the line-of-sight direction.

For the magnetic field strength, the line-of-sight component was used in the analysis, 
and the inclination was analyzed using the inclination of the magnetic field lines 
from the local zenith of the photosphere. 
The field inclination was transformed from the line-of-sight coordinate system 
to the local coordinate system. 
We adopted the angle at which the field inclination becomes small relative to 
the local zenith due to the $180^\circ$ ambiguity resolution 
because the magnetic field lines are expected to continuously tilt from the center of the pore to the periphery.


\subsection{IRIS observation and analysis
\label{IRISanalysis}}
IRIS was observed with the slit (width $0.166''$) for spectroscopy and
with the slit jaw imager (SJI), which measures the radiation intensity 
by imaging observations. 
The time variation of the Doppler velocity was derived 
from spectroscopic observations of the $\rm Mg\ I\hspace{-.1em}I\ k$ line. 
The spectral sampling is $12.72\rm m\AA$, and the time resolution is 27 s. 
The slit scanned a $7''$ width in 8 steps at $1''$ intervals. 
The SJI was acquired every 14 s at four wavelengths: $\rm C\ I\hspace{-.1em}I\ 1330\AA$, $\rm Si\ I\hspace{-.1em}V\ 1400\AA$, $\rm Mg\ I\hspace{-.1em}I\ k\ 2796\AA$, and $\rm Mg\ I\hspace{-.1em}I\ wing\ 2830\AA$. The SJI enabled the identification of the exact position of the slit. We used the level-2 data created with the instrumental calibration including the dark current subtraction, flat field, and geometrical corrections (\citealt{DePontieu2014}). 

The wing of the $\rm Mg\ I\hspace{-.1em}I\ k$ line provides 
information about the lower part of the chromosphere, whereas the core is 
sensitive to the upper part of the chromosphere 
(\citealt{Leenaarts2013a, Leenaarts2013b}, \citealt{Pereira2013}). 
The $\rm Mg\ I\hspace{-.1em}I\ k$ line is typically observed in double 
peaks in the quiet-Sun and single peaks in sunspots. 
The plage may show different shapes, with brighter, wider wings, 
single peaks, or inverted cores (\citealt{Morrill2001}). 
The depth of the core inversion is inversely proportional to the magnetic field strength. 
This is because the $\rm Mg\ I\hspace{-.1em}I\ k$ line is a very abundant elemental resonance line; 
therefore, the line core is optically thick and can form at relatively low densities 
in nonlocal thermodynamic equilibrium (\citealt{Carlsson2015}, \citealt{Schmit2015}).

\begin{figure}[h]
\centering
\subfigure[wing]{
 \includegraphics[keepaspectratio, scale=0.8]{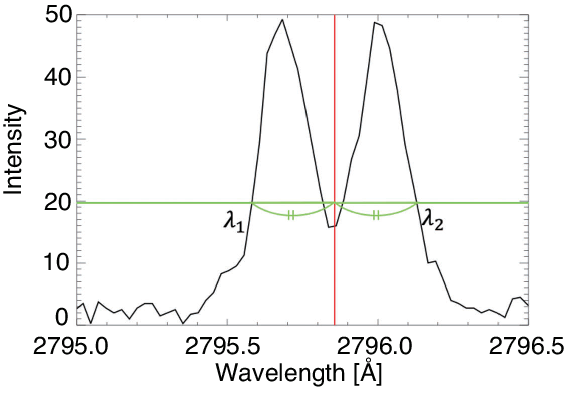}
 \label{wing}}\\
 \subfigure[core]{
 \includegraphics[keepaspectratio, scale=0.8]{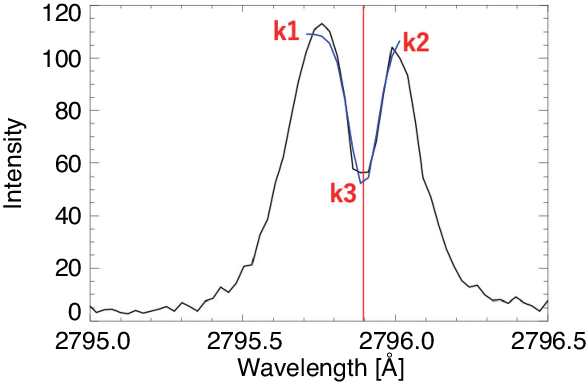}
 \label{core}}\\
 \subfigure[core (single peak)]{
 \includegraphics[keepaspectratio, scale=0.8]{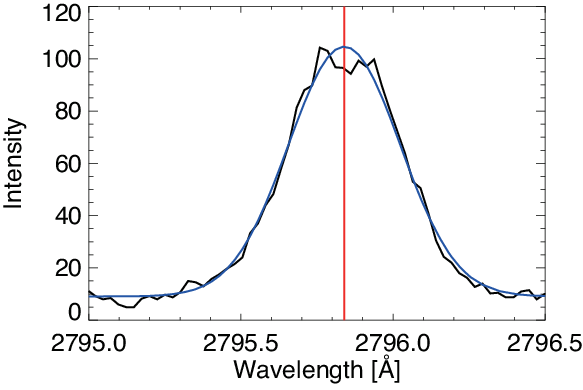}
 \label{core_peak}}\\
 \caption{Analysis for $\rm Mg\ I\hspace{-.1em}I\ k$ 
     to derive line-of-sight velocity. (a) The bisector at 40\%
     of the peak intensity for velocity at the lower chromosphere.
     (b) The Gaussian fitting to a core inverted profile 
      for velocity at the upper chromosphere. 
    (c) The Gaussian fitting to the whole line profile is used instead
      if the core inverted profile is tiny.
 }\label{Mg2k}
\end{figure}

The bisector method was used to derive the wavelength of the line center at the intensity level
of the line wings; it is the midpoint of $\lambda_1$ and $\lambda_2$ defined at 40\% of 
the peak intensity of $\rm Mg\ I\hspace{-.1em}I\ k$,  as shown in Figure \ref{wing}
(\citealt{Graham2015}). 
The reference wavelength for the zero Doppler velocity is derived by averaging 
the line profiles averaged in the spatial and temporal directions for all pixels 
on each day. 
The quiet Sun is expected to have intrinsic velocity of zero or less than 
a few $\rm km/s$. However, as we did not obtain data for the quiet Sun in 
this observation, we assume that the velocity is zero, averaged over the observation 
region in the spatial and temporal directions. 
The Doppler velocity was obtained from the deviation of the wavelength from the reference wavelength.

The wavelength of the line core was defined by detecting three extremes $k_1,k_2,k_3$ ($k_3$ is 
the minimum value between $k_1$ and $k_2$) and applying Gaussian fitting in the range from
 $k_1$ to $k_2$ (Figure \ref{core}). 
In the plage region, $\rm Mg\ I\hspace{-.1em}I\ k$ may be observed as 
a single peak with a tiny core inverted profile, as shown in Figure \ref{core_peak}. 
In such cases, profiles where (intensity decrease at $k_3$)/(average of the intensities with 
respect to $k_1$ and $k_2$) is less than 15\% are considered to be 
at single peaks, and the wavelength was derived with the Gaussian fit to the whole line profile instead.

\subsection{Co-alignment of spectral lines}
The Hinode SOT/SP and IRIS fields of view were aligned to identify the IRIS slit positions on the SP field of view. The SJI images of $\rm Mg\ I\hspace{-.1em}I\ wing\ (2832\AA)$, which are dominantly photospheric, 
were used for the alignment. 
Each image was cross-correlated with the SP map taken at the closest time, 
providing the relative displacement between them. 

\begin{figure}[h]
\centering
\includegraphics[keepaspectratio, scale=0.5]{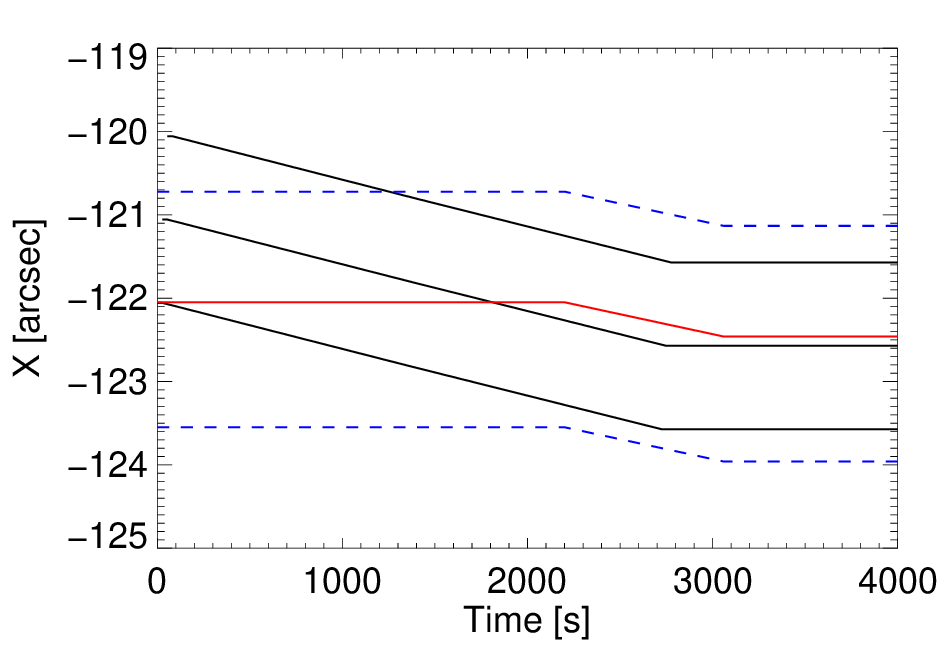}
\caption{SP and IRIS fields of view in the x-direction as a function of time. Black line: the slit position of the three IRIS slits located within the SP field of view. Blue dotted lines: the ends of SP field of view. Red line: the center of SP field of view. The figure is for the data captured on February 10, 2018.}
\label{position}
\end{figure}

Figure \ref{position} shows how the SP and IRIS fields of view were drifted in the X, i.e., E-W direction
on the solar surface as a function of the time. 
At least two IRIS slits are included in SP field of view. 
IRIS slits in the SP field-of-view were selected, and SP pixel data overlapping 
with the selected slits were used for analysis. 
The positions in the N-S direction were also co-aligned in the same manner. 

\section{Results} \label{sec:results}

\begin{figure}[h]
\centering
\includegraphics[keepaspectratio, scale=0.5]{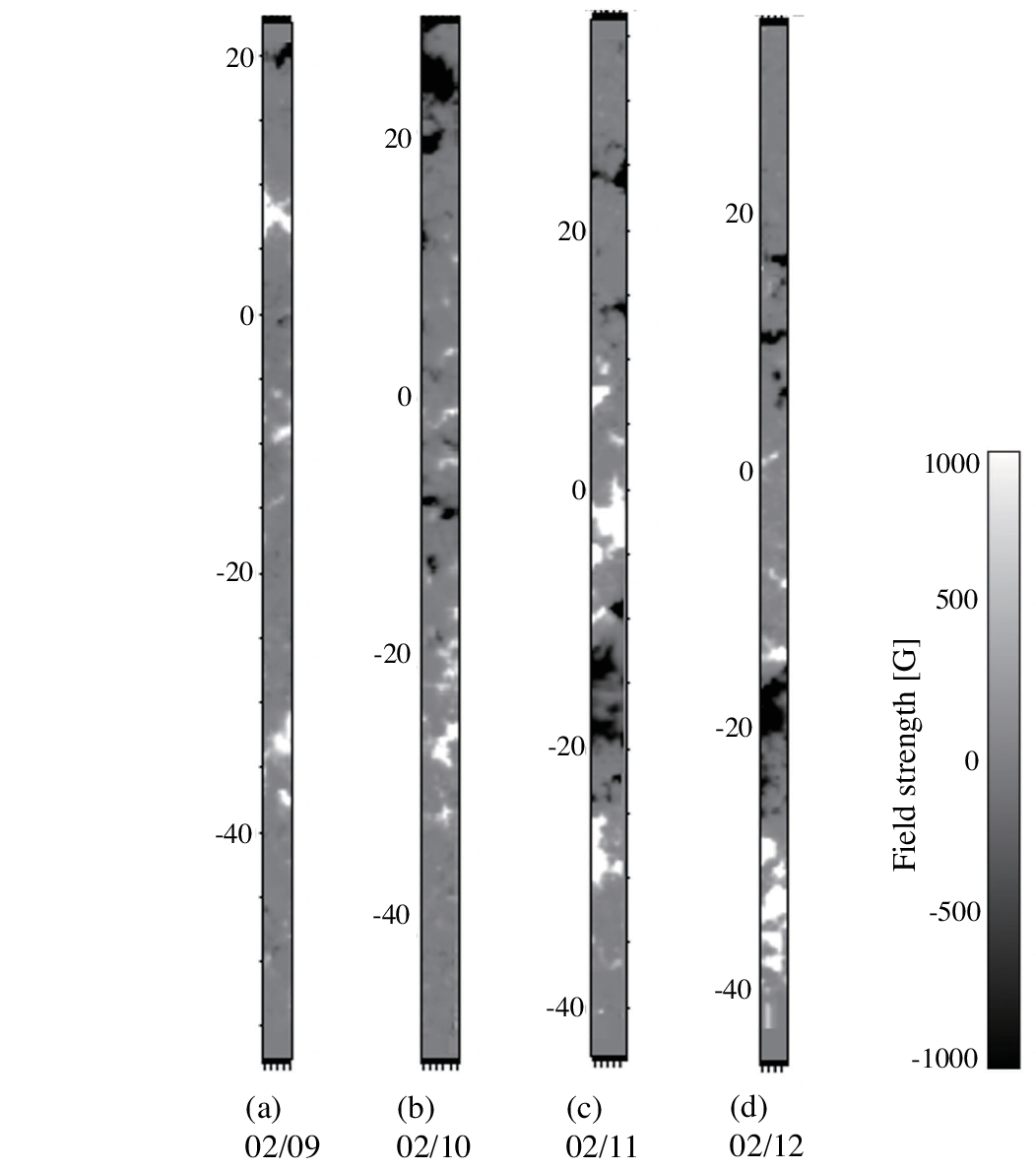}
\caption{ LOS magnetic field strength. 
 The unit on the vertical axis is arcsec.}
 \label{magnetogram}
\end{figure}

\begin{figure}[h]
\centering
 \includegraphics[keepaspectratio, scale=0.5]{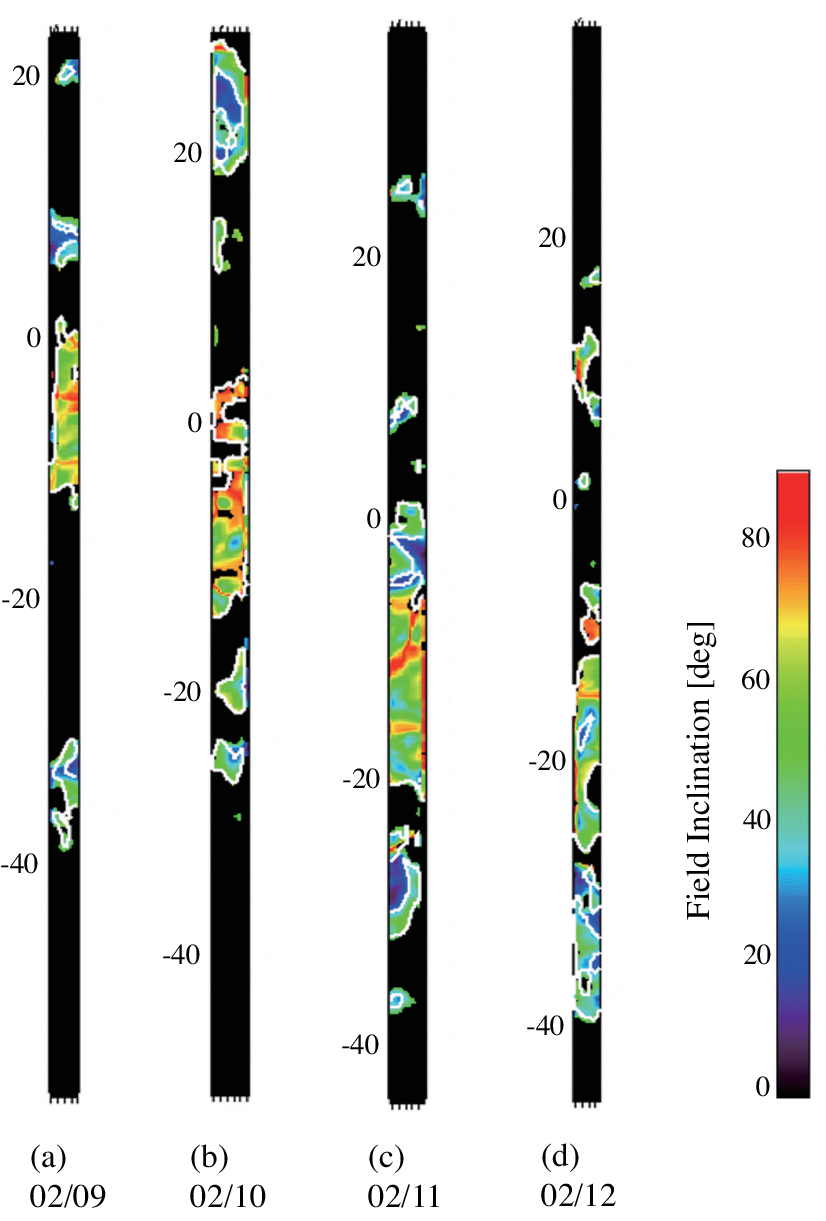}
 \caption{Inclination of magnetic field lines.
  The unit on the vertical axis is arcsec.}
  \label{inclination}
\end{figure}

The line-of-sight magnetic field strength and the inclination of the magnetic field in the observation regions are shown in Figures \ref{magnetogram} and \ref{inclination}. 
The inclination of the magnetic field, denoted by $\theta$, 
is defined in the local frame coordinate, i.e., $0^\circ$ in the normal direction of the solar surface and $90^\circ$ in the horizontal direction. From the negative polarity of $y=20''$ -- $25''$ in Figure \ref{inclination}, 
the magnetic field lines are continuously inclined from the center of the pore (a small sunspot with no penumbra), where the magnetic field strength is relatively strong, to the surroundings. Pixels with the degree of polarization of less than 1\%, which are considered unsuitable for Stokes profile inversion with good accuracy, were not used in the analysis.

\begin{figure}[h]
\centering
 \subfigure[In the photosphere derived from the SP data]{
  \includegraphics[keepaspectratio, scale=0.6]{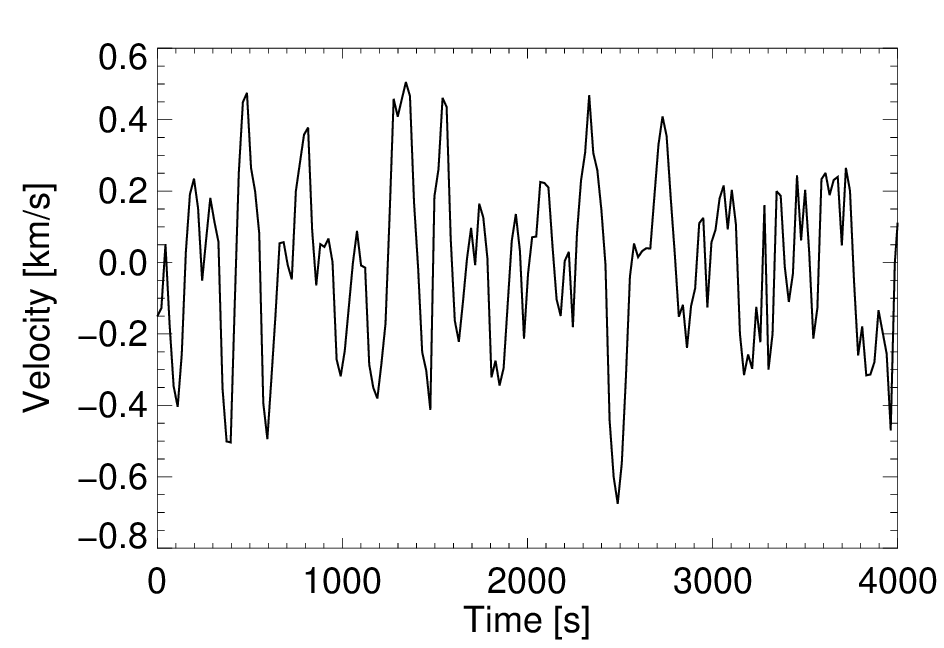}}\\
  \subfigure[In the lower chromosphere derived from IRIS $\rm Mg\ I\hspace{-.1em}I\ k$ wing]{
  \includegraphics[keepaspectratio, scale=0.6]{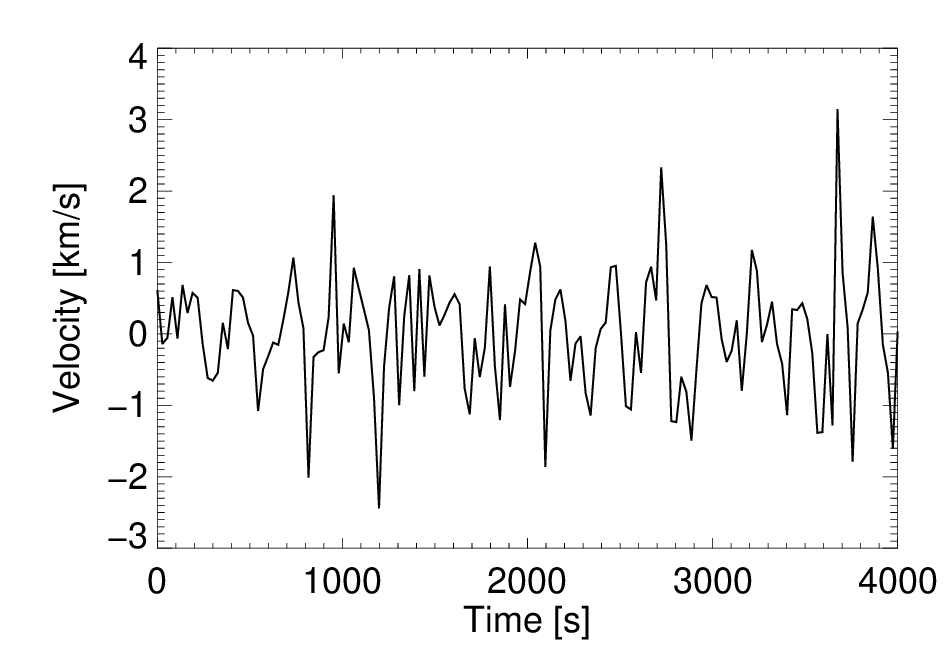}}\\
  \subfigure[In the upper chromosphere derived from IRIS $\rm Mg\ I\hspace{-.1em}I\ k$ core]{
  \includegraphics[keepaspectratio, scale=0.6]{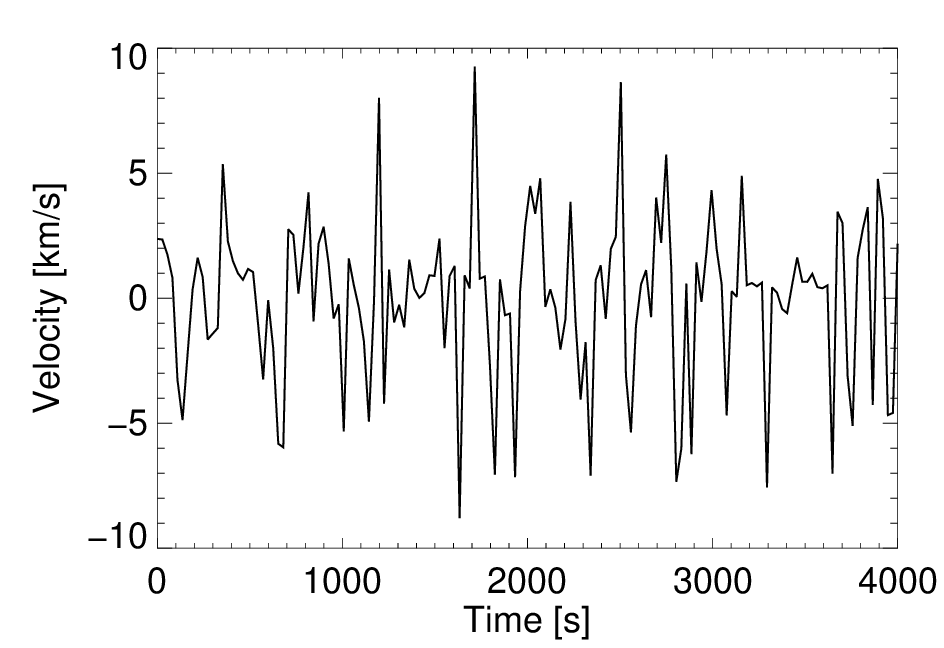}}\\
\caption{Time series of the Doppler velocity observed in a pixel region 
  on February 10, 2018.}\label{wave}
\end{figure}

Figure \ref{wave} shows an example of the time series of the Doppler velocities observed in a pixel on February 10, 2018. The pixel has a weak magnetic field strength (353 G) and tilted magnetic field lines ($\theta=54^\circ$). The amplitudes in each layer are observed to be within 0.8 km/s in the photosphere, approximately 2 km/s in the lower chromosphere, and up to 10 km/s in the upper chromosphere. The amplitude of the waves is larger in the upper layers. This is because the density decreases in the upper layers. We derived the power distribution by Fourier transforming each wave because determining the phase difference from the waves observed in each layer is difficult.

\subsection{Dependence of power spectrum on inclination of magnetic field lines}

\begin{figure}[h]
\centering
\subfigure[In the photosphere derived from the SP data]{
 \includegraphics[keepaspectratio, scale=0.6]{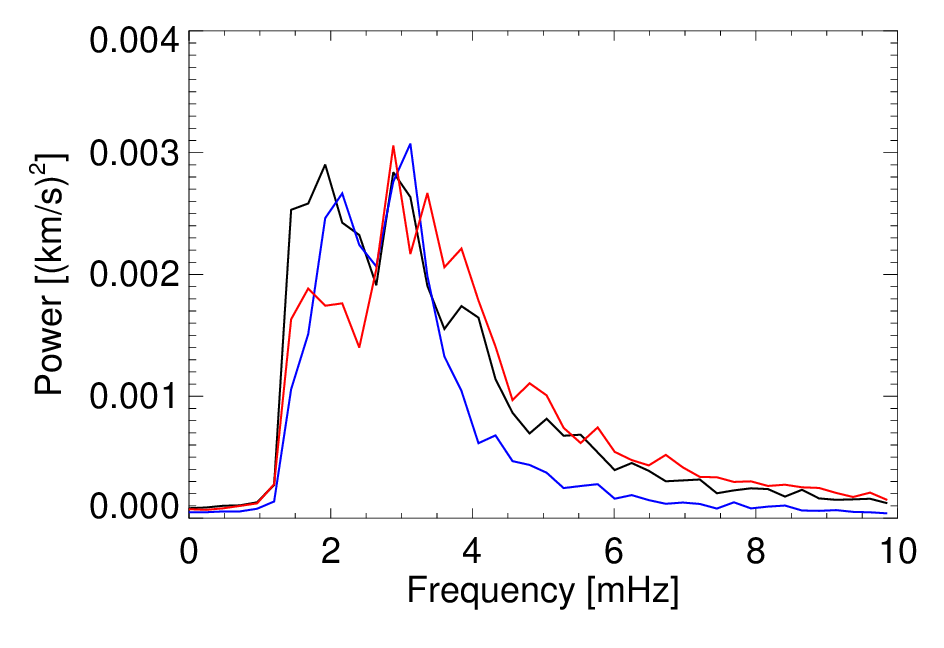}
 \label{power_SP}}\\
 \subfigure[In the lower chromosphere derived from IRIS $\rm Mg\ I\hspace{-.1em}I\ k$ wing]{
 \includegraphics[keepaspectratio, scale=0.6]{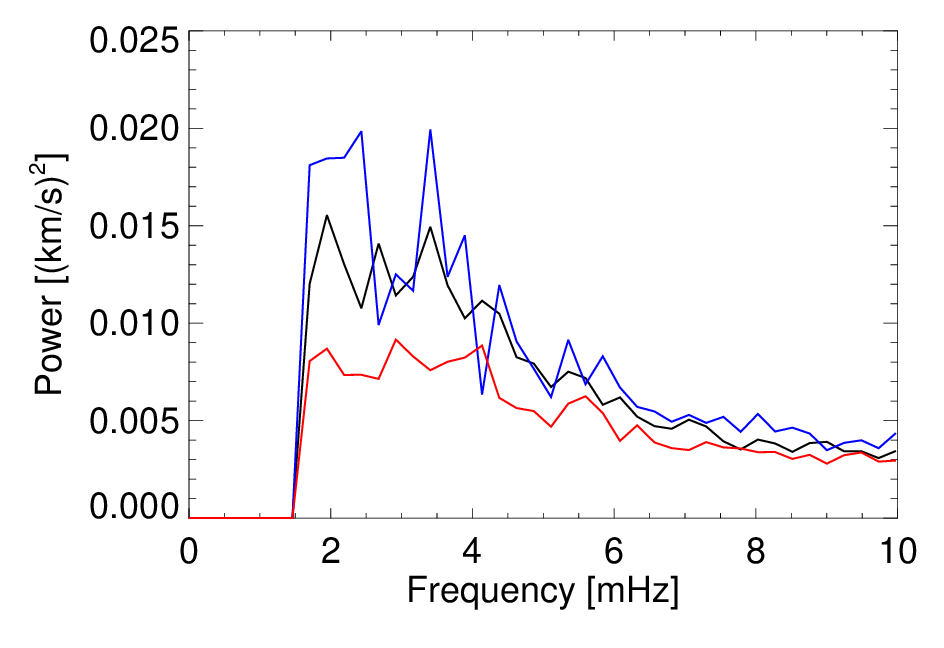}
 \label{power_wing}}\\
 \subfigure[In the upper chromosphere derived from IRIS $\rm Mg\ I\hspace{-.1em}I\ k$ core]{
 \includegraphics[keepaspectratio, scale=0.6]{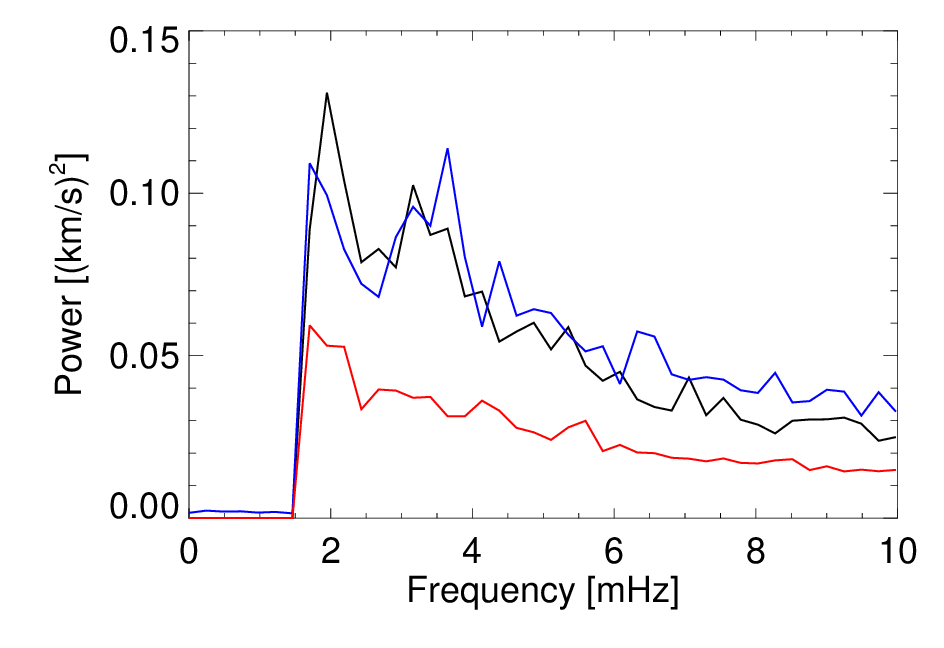}
 \label{power_core}}\\
 \caption{Power distribution as a function of the frequency. The inclination of the magnetic field lines; blue lines show $0^\circ<\theta<30^\circ$, black lines show $30^\circ<\theta<50^\circ$, red lines show $50^\circ<\theta<90^\circ$.}\label{power}
\end{figure}

The Fourier transform was applied to the time profiles to derive the power distribution 
as a function of the frequency. 
Figure \ref{power} shows the power spectrum for the three sets of time profiles:
(a) in the photosphere derived from the SP data, 
(b) in the lower chromosphere derived from the wing of the $\rm Mg\ I\hspace{-.1em}I\ k$ line in IRIS, 
and (c) in the upper chromosphere derived from the core. 
The power spectra are the averages of the five-day data.  
The inclination of the magnetic field lines $\theta$ is classified into $0^\circ<\theta<30^\circ$, $30^\circ<\theta<50^\circ$, and $50^\circ<\theta<90^\circ$. For all data, a subsonic filter (\citealt{Title1989}, \citealt{Oba2017}) was applied before the Fourier transfer to remove the frequency component caused by the gas convection at the photosphere. The power distribution below 1.5 mHz was removed because the gas convection is below 2 mHz. 
Figure \ref{power_SP} shows that the power distribution in photosphere is 
enhanced 
at 3 mHz. 
This may be due to the observation of the 5-min oscillations (3.3 mHz). 
Below 5 mHz, there is no difference in the power with respect to the inclination of the magnetic field lines, while above 5 mHz, the power of $\theta>30^\circ$ is slightly larger than that of $\theta<30^\circ$.
 As shown in Figure \ref{power_wing}, in the lower chromosphere, 
 the power of $\theta>50^\circ$ is smaller than the others
 below 5 mHz, and there is no difference with respect to the inclination of the magnetic field lines
 above 5 mHz. 
 Figure \ref{power_core} shows that the power of $\theta>50^\circ$ in the upper part 
 of the chromosphere is smaller than the others for all frequencies.


\subsection{Energy dissipation in the photosphere and chromosphere
\label{EnergyDissipation}}

The energy flux $F$ for the three heights at the photosphere, lower and upper chromosphere, can be estimated using the following equation (cf. \citealt{Sobotka2016})
\begin{equation}
\label{energyflux_Sobotka}
F=\int_{\nu_{ac}}^{\nu_{max}} \rho P_v(\nu)v_{g}(\nu) d\nu
\end{equation}
where $\rho$ is the density, $P_v(\nu)$ is the power spectrum of the Doppler velocity obtained from the Fourier transform,
and $v_{g}$ is the group velocity for propagating waves. 
The group velocity for 
waves propagating along inclined magnetic field lines is 
given by the following equation using the cutoff frequency $\nu_{ac}$ obtained from Equation (\ref{Eqcutoff}) and the sound speed $c_s=\sqrt{\gamma p/\rho}$,
\begin{equation}
\label{vgr}
v_{g}=c_s\sqrt{1-(\nu_{ac}/\nu)^2}.
\end{equation}
 Here the acoustic slow wave is essentially field-guided \citep{Schunker2006} and we therefore multiply 
 the geometric factor $\cos \theta$ in the calculation of the energy flux, 
 in order to derive the energy flux which 
 the acoustic waves bring in
 the vertical direction upward from the solar surface.
The density $\rho$ and temperature $T$ at each height are taken from the atmospheric model VAL-F (\citealt{Vernazza1981}), which is an atmospheric model for bright inter-network regions.

The energy dissipation at the photosphere was estimated from the difference between the lower chromosphere and the photosphere, while the energy dissipation at the chromosphere was estimated from the difference between the upper and lower chromospheres. The error is obtained from the following equation using the power spectrum error $\sigma_{P_v}$.
\begin{equation}
\label{sigma}
\sigma_F=\sqrt{\sum_v(\rho v_{g})^2(\sigma_{P_v})^2}
\end{equation}

\subsubsection{Dependence on the inclination of magnetic field
\label{Secenergy}}

\begin{figure}[h]
\centering
\includegraphics[keepaspectratio, scale=0.9]{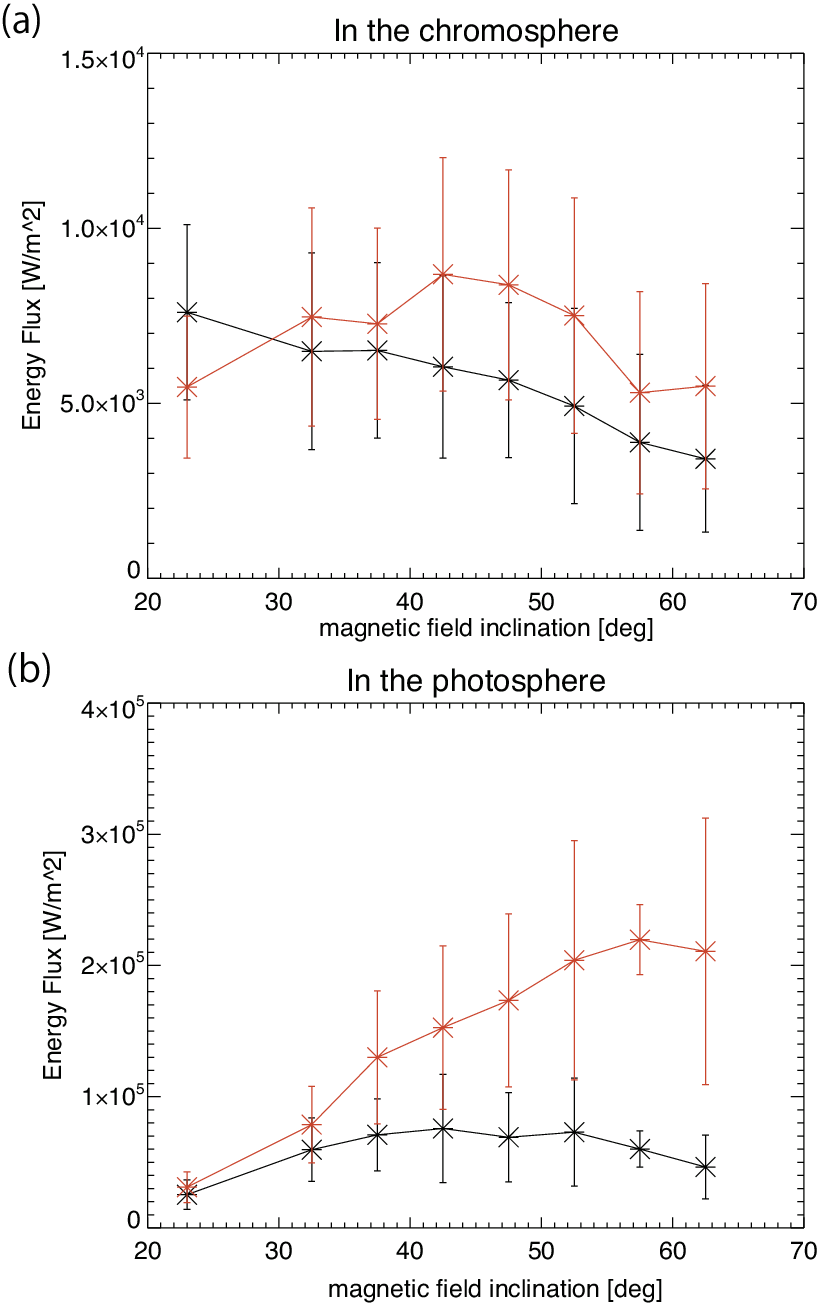}
\caption{
The average amount of energy dissipated
(a) in the chromosphere and (b) in the photosphere
as a function of the inclination of the magnetic field lines. 
Red line: low frequency (3--6 mHz), Black line: high frequency (6--10 mHz).
}\label{flux}
\end{figure}

The 
average
amounts of energy dissipated in the photosphere and chromosphere 
are shown in Figure \ref{flux}(b) and Figure \ref{flux}(a), respectively, 
as a function of the inclination of the magnetic field. 
The inclination is measured at the photosphere (Figure~\ref{inclination}). 
The red line represents the amount of energy dissipated by low frequency (3--6 mHz) component, 
while the 
black line represents the amount of 
energy dissipated by high frequency (6--10 mHz) component. 
The inclination of the magnetic field lines $\theta$ is classified into $\theta<30^\circ$, 
$5^\circ$ each for $30^\circ<\theta<60^\circ$,
and $\theta>60^\circ$.

In the photosphere, the low-frequency waves make more contributions to the energy dissipation. 
The energy by low-frequency waves increases monotonically with the field inclination, 
reaching the amount of the energy dissipation exceeding $2 \times 10^{5}$ W/m$^{2}$ at the field inclination 
around $50^\circ-60^\circ$. The heating contributions of high-frequency waves to the photosphere are
almost constant, less dependent of the field inclination, and the amount is around $(2 - 6) \times 10^{4}$ W/m$^{2}$. 

The dependency of the energy dissipation in the chromosphere is different from that observed in the photosphere. 
The energy dissipated in the chromosphere by the high-frequency waves decreases monotonically
as a function of the field inclination; $8 \times 10^{3}$ W/m$^{2}$ in the inclination smaller than $30^\circ$
and $3 \times 10^{3}$ W/m$^{2}$ in the inclination larger than $60^\circ$.
The energy by low-frequency waves increases with the field inclination in the range below $40^\circ$,
whereas the energy decreases in the range above $40^\circ$. 
The maximum energy dissipation is $9 \times 10^{3}$ W/m$^{-2}$  on average at the field inclination of $40^\circ$,
which may be slightly lower than the energy required for heating in the chromosphere. 
\cite{Jefferies2006} demonstrated that low-frequency waves propagate to the upper layers 
in the region where the magnetic field lines are inclined more than $30^\circ$, but Figure \ref{flux}(a)
shows that the energy may not be carried efficiently to the chromospheric height 
when the magnetic field lines are more inclined beyond $40^\circ$.  

\subsubsection{Dependence on the inclination of magnetic field lines and field strength}

\begin{figure}[h]
\centering  
\includegraphics[keepaspectratio, scale=0.75]{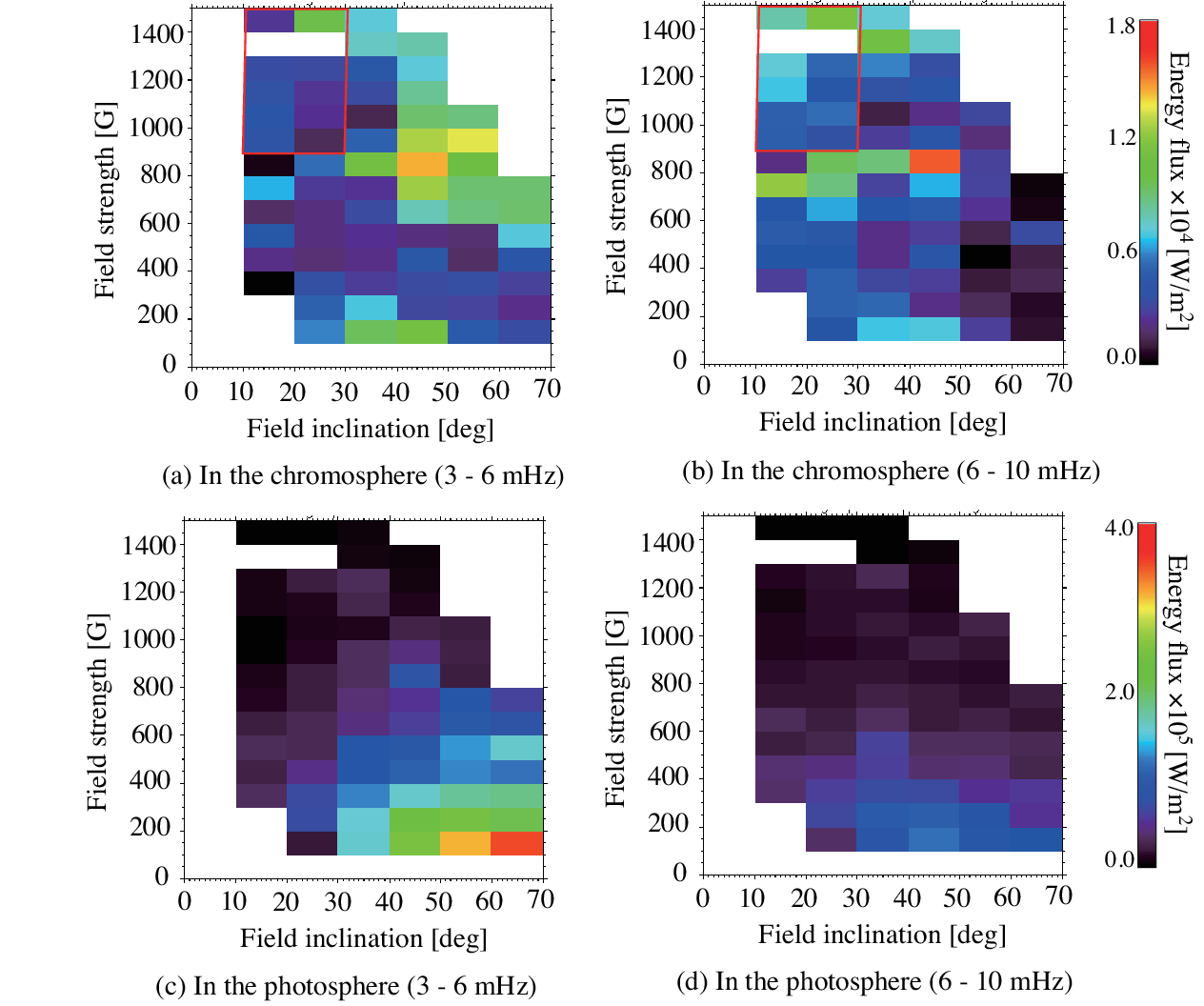}
\caption{
Amount of the dissipated energy as a function of the inclination of the magnetic field lines 
in the horizontal axis and the magnetic field strength in the vertical axis. 
(a) and (b) are the energy dissipated in the chromosphere by the low-frequency component (3-6 mHz) and
the high-frequency component (6-10 mHz), respectively. (c) and (d) are the same but for the photosphere.
The red box in (a) and (b) gives the condition where the line profile of $\rm Mg\ I\hspace{-.1em}I\ k$ line is 
a single peak with only a tiny core inverted profile. The velocity for the upper chromosphere was derived
by a whole profile fitting instead of the fitting to the core inverted profile as described in section \ref{IRISanalysis}.  
The values are acquired by averaging the data for four days. 
}
\label{fdegmag}
\end{figure}

Figure \ref{fdegmag} shows the energy dissipated in the chromosphere and photosphere with respect to the magnetic field strength (line of sight) and inclination. The field inclination is in $10^\circ$ increments
in the horizontal axis,
and the field strength $|B|$ is in $\rm 100\ G$ increments
in the vertical axis.
The energy dissipation values are the average of the data in four days.

For the amount of the dissipated energy estimated in the chromosphere, 
the low-frequency component shows an enhancement when
the magnetic field lines with the magnitude higher than 600 G are inclined
more than $40^\circ$ (Figure \ref{fdegmag}(a)),
while no 
apparent
dependence on the inclination angle
is
observed for the high-frequency component (Figure \ref{fdegmag}(b)). 
The low-frequency and high-frequency components show less enhancements of 
the dissipated energy in the weak field pixels, which is different from behaviors in the photosphere
as seen in Figure~\ref{flux}(c). 
The spectral profiles with single or small amplitude peaks (Figure \ref{core_peak}) are observed mainly in the region 
marked by red box in Figures \ref{fdegmag}(a) and (b),
indicating that they do not considerably affect the overall trend.
If we focus on the energy dissipation in the photosphere, 
the energy dissipation of the low-frequency component 
is enhanced in the magnetic field below 600 G (Figure \ref{fdegmag}(c)).
The enhanced energy dissipation is also more significant as the field inclination increases. 
For the high-frequency component
(Figure \ref{fdegmag}(d)), no significant enhancements are observed but the dissipated energy
shows weak dependence on the field inclination,
Also, the amount of energy dissipation is slightly enhanced in the magnetic field below 600 G.
It is different from that observed in the low-frequency component.

The maximum 
magnitude of the
energy dissipation by the low frequency component in the chromosphere 
is $\rm 1.4\times10^4\ W/m^2$ (The value averaged over four days) 
at $40^\circ<\theta<60^\circ$ and $\rm 800~G<|\it B\rm|<1,000~G$.
The dissipated energy may exceed $\rm 1\times10^4\ W/m^2$ 
when the magnetic field with the strength higher than 600 G is inclined
by 40 degree or more from the solar surface.
This 
may satisfy
the heating required for the chromosphere. The maximum energy dissipation in the photosphere 
is $\rm 3\times10^5\ W/m^2$ 
(averaged over four days), which are distributed in weak ($<400$ G) and inclined
($> 50^\circ$) field pixels.

\section{Discussions}

\subsection{Interpretations of the results}
\begin{figure}[h]
\centering
\includegraphics[keepaspectratio, scale=0.7]{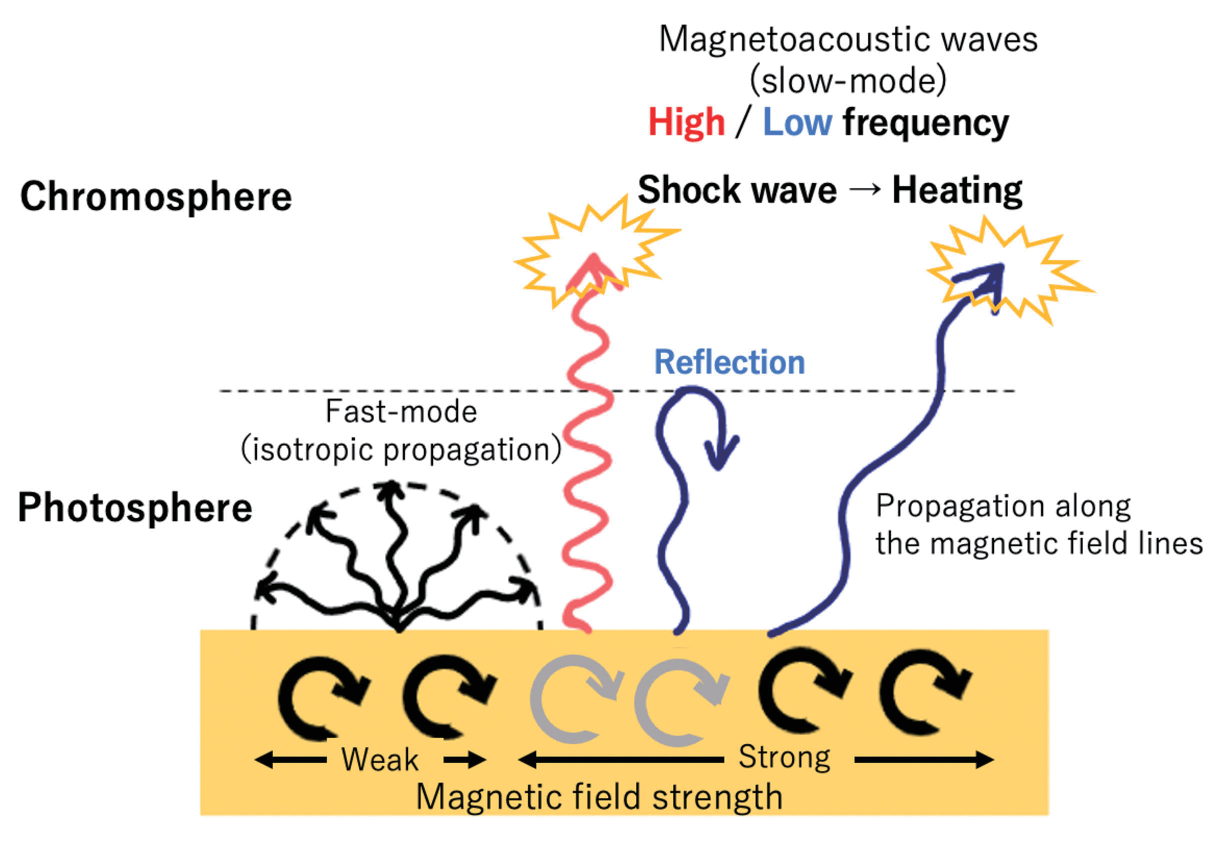}
\caption{Waves inferred from results (schematic)} \label{wavemoshi}
\end{figure}

As shown in Figure \ref{fdegmag},  
the energy dissipation of the low-frequency component is more significant in the photosphere 
when the magnetic field is weaker than 400 G,
while the chromosphere shows more energy dissipation by the low-frequency component 
when the magnetic field lines with the strength 
higher than 600 G are inclined more than $40^\circ$ from the solar surface. 
On the photosphere, a large number of acoustic oscillation modes (p-modes) are mainly excited
with periods of about 3-15 min (frequencies of about 1-5 mHz) and they are observed
as the low-frequency component in weak magnetic field regions. 
Gas convection, which is the
generator
of magnetoacoustic waves at the photosphere, becomes weaker in the strong magnetic fields, making it harder for waves to be generated. 
The weak magnetic field region may contain isotropically propagating waves (fast mode of magnetoacoustic waves), which would dissipate before propagating to the chromosphere. 
As the amount of energy dissipation depends on the inclination of the magnetic field lines 
in the low-frequency component, the reduced effective gravity 
on the inclined magnetic field lines makes the cutoff frequency lower, 
allowing for the low-frequency component of the waves to propagate along the magnetic field lines and to dissipate in the chromosphere. This interpretation suggests that the slow mode waves propagating along the magnetic field lines contribute significantly to the chromospheric heating. The left side of Figure \ref{wavemoshi} shows that fast-mode waves propagate isotropically in the weak magnetic field, 
and the center shows that high-frequency component can propagate upward into the chromosphere
while low-frequency component is reflected below the cutoff frequency.  
The right side shows that low-frequency component can propagate upward 
because of lower cutoff frequency caused by the inclination of the magnetic field lines.

We have assumed here that the traveling waves generated by the photospheric convection 
heat the upper atmosphere; however, our analysis cannot provide the phase relations  
in the temporal profiles of velocities observed at each layer. 
Thus, we cannot give a definitive conclusion that the amount of energy 
obtained in this study is due to the dissipation of the traveling waves and their transformation into thermal energy. A part of velocity fluctuations may come from the waves reflected at the upper chromosphere and go downwards. There are two reasons why determining the phase relations 
in velocity fluctuations measured at different heights is difficult; 
(1) when observing waves propagating along inclined magnetic field lines, 
the line-of-sight direction in the two measurements may not give the information 
on the same field line so that the same waves may not be captured at the two layers; 
(2) the chromosphere is full of plasma flows, making it difficult to distinguish 
the waves propagating from the photosphere. Figure \ref{power_wing} and (c) showed 
that the power in the magnetic field lines with $\theta>50^\circ$ is smaller than the others. 
This reduction may be partially explained by the Doppler velocity measurements as the line-of-sight component of waves propagating along the magnetic field lines.

For inclined magnetic field lines, the line-of-sight measurements at two layers 
may not derive the energy flux on the same magnetic field lines. As a trial, we shifted the observation position in the lower chromosphere (IRIS wing) by $1''$ -- $2''$ in the east-west direction from the original observation position and compared the power spectrum with that at the photosphere. The result demonstrated that the power spectrum in the lower chromosphere was larger than that in the photosphere for the low-frequency component depending on the observation position in the lower chromosphere. The data used herein have an extremely narrow field of view to increase the cadence, leading to difficulty in estimating the orientation of the inclined field lines. The 180-degree ambiguity resolution requires the spatial distribution of the field for a much wider field of view. To investigate the energy dissipation in more detail, tracing the magnetic field lines from the photosphere to the chromosphere is necessary. In the near future, the SUNRISE-3 experiment, that is,
the third flight in series of the SUNRISE balloon-borne stratospheric observatory
with a 1-m solar telescope \citep{Barthol2011}, is 
under planning for re-flight in 2024 and it will provide high-cadence series of 
spectro-polarimetric data for much wider field of view and with
coverage from the photosphere to the chromosphere.
Particularly, the Sunrise Chromospheric Infrared spectroPolarimeter \cite[SCIP; ][]{Katsukawa2020}
on the SUNRISE-3 experiment will be able to provide some spectral lines seamlessly covering from the photosphere
through the middle chromosphere \citep{QuinteroNoda2017}. 
Moreover, the focal plane instruments equipped to the ground-based 4-m 
Daniel K. Inouye Solar Telescope (DKIST), a facility of the National Solar Observatory (NSO), 
will provide similar series of spectro-polarimetric data.

The gas convection at the photosphere is a source of waves. The turbulent motions excite movement and deformation at the foot of the magnetic field lines. If the opposite polarity flux exists next to the field lines, they may approach each other and cause magnetic reconnection, 
i.e., nanoflares (\citealt{Parker1972}), which may also generate another type of waves. 
Such a wave can contribute to velocity fluctuations. However, 
imagining that such a wave is included as the dominant source may be difficult because the energy dissipation of the low-frequency component increases with the inclination of the magnetic field lines, 
not only in the photosphere but also in the chromosphere at least with the inclined
magnetic field up to 40 degree. 
Our results have been interpreted with slow mode waves as the most possible mechanism, but velocity fluctuations can also be created by Alfv\'{e}n waves, which are a candidate for the heating of the corona. However, owing to the incompressible nature, Alfv\'{e}n waves are unlikely to develop to shock waves in the chromosphere and may contribute little to the chromospheric heating.

\subsection{Formation height and density's assumption}

As described in Section \ref{EnergyDissipation}, we used densities from the VAL-F atmospheric 
model to derive the energy flux in Figure \ref{flux}. 
The VAL-F model is an atmospheric model for the bright inter-network 
that reproduces an atmosphere closer to the plage we have studied. 
Here we discuss at which height the three layers analyzed, i.e., 
the photosphere, the lower chromosphere, and the upper chromosphere, are located from 
the solar surface and evaluate how the estimated energy is sensitive to 
the assumption of density used in the calculation.  
The SP data is mostly sensitive to approximately 300 km 
from the solar surface, based on the formation height of the $\rm Fe\ I\ 6301.5\ \AA$ line,
and thus, velocities from the SP were considered for the photospheric energy flux.

\begin{figure}[h]
\centering
\includegraphics[keepaspectratio, scale=0.6]{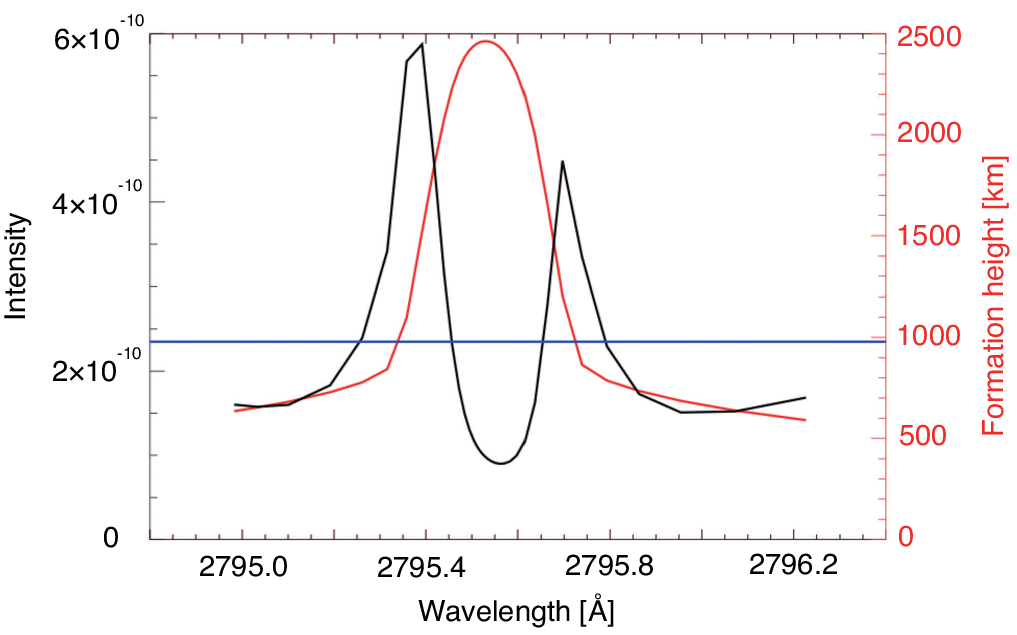}
\caption{Spectral profile of the $\rm Mg\ I\hspace{-.1em}I\ k$ line averaged over the field of view 
(black), and the height of the formation at each wavelength position of the line (red).} \label{height}
\end{figure}

On the derivation of Doppler velocities at the lower and upper chromospheric layers, we
used the wing and core profiles in the $\rm Mg\ I\hspace{-.1em}I\ k$ line, respectively.
To evaluate the height where the wing and core portions of the line are formed,
we examined a publicly available simulation data cube
from a realistic simulation of an enhanced network region \citep{Carlsson2016}. 
This simulation used the radiation magnetohydrodynamics code {\em Bifrost}
\citep{Gudiksen2011} to simulate the magnetic structure and dynamics of 
the outer atmosphere with a magnetic field topology characterized by
two dominant opposite polarity islands separated by 8 Mm. 
The simulated magnetic field topology may be similar to what is observed in
our plage regions located between two opposite polarity sunspots, although 
the spatial scale of the magnetic bipolar structure in the simulation is much smaller
than that of our region.  
Nevertheless,  we synthesized spectral profiles of the $\rm Mg\ I\hspace{-.1em}I\ k$ line
in a snapshot from this simulation for an enhanced network region by 
non-LTE radiative transfer computations with {\em RH} code \citep{Uitenbroek2001} 
and the method described in detail by \cite{Leenaarts2013b} and \cite{Hasegawa2021}. 
This computation enables us to study the heights where the wing and core of 
the $\rm Mg\ I\hspace{-.1em}I\ k$ line are formed. 
Figure \ref{height} shows the spectral profile averaged over the field of view 
in a snapshot and its height of the formation at each wavelength position in the central portion of the line.
The wing portion for the lower chromosphere is formed at the height approximately 1,000 km,
whereas 2,000--2,500 km for the line core that was fitted for the upper chromosphere.
These values are only approximate because they are 
derived from the averaged spectral profile and are not 
identical to the actual atmosphere observed herein. 
As the chromosphere is thinner in the active region than in the quiet region, 
the density at 2,000 km (10,000 K) in VAL-F was assumed for the density in 
the upper chromosphere.   
When the formation height of the line core 
changes from 2,000 km (used in the results presented so far) to 1,800 km and 2,400 km 
in the VAL-F model, 
such difference in the formation height of the line core
provides a very minor change 
(less than 2 percent) 
in the evaluated energy dissipation in the chromosphere.




\section{Summary}

In this study, we quantitatively estimated the magnetic field structure and
derived how
the energy dissipations at the photosphere and chromosphere
depend on the properties of magnetic fields
in the plage using the simultaneous observations 
by Hinode and IRIS. 
The results show that the energy dissipation by low-frequency waves at 3--6 mHz 
increases at the chromosphere 
in the range of the magnetic field inclination below 40 degree, and
it begins to decrease when the field is more inclined beyond 40 degree.
This is considered to be due to the propagation of low-frequency waves into the chromosphere.
The effective gravity on the magnetic field lines decreases due to the tilt of the magnetic field 
lines, resulting in decreasing cutoff frequency. 
We interpreted 
that the waves observed in the chromosphere are mostly slow-mode waves propagating 
along the magnetic field lines. 
The low-frequency waves may be able to bring the energy in the vertical direction, although
too highly inclined field may suppress the transport of energy in the vertical direction.
The plage is observed to be bright in the chromosphere and is considered to 
be important for chromospheric heating. 
The previous studies, however, have not quantitatively described 
the energy dissipation above the plage as far as their authors know. 
Few studies have quantitatively utilized the magnetic field structure derived in the photosphere. 
The energy flux obtained by previous studies is less than the heating rate required for the chromosphere (\citealt{Sobotka2016}). In this study, we found that the low-frequency component in 3--6 mHz 
can dissipate
more than $10^4\ \rm W/m^2$ of energy in the chromosphere
in only a specific configuration of the magnetic field lines.
Furthermore, the amount of energy dissipated in the photosphere increases with decreasing magnetic 
field strength at the photosphere, 
whereas the amount of energy dissipated in the chromosphere is 
more significant in the magnetic field strength higher than 600 G.



\vspace*{22pt}
\begin{acknowledgments}
The authors thank the anonymous referee for pointing out a flaw in the energy assessment in draft manuscript. 
Hinode is a Japanese mission developed and launched by ISAS/JAXA, with NAOJ as domestic partner and NASA and STEC (UK) as international partners. It is operated by these agencies in co-operation with ESA and NSC (Norway). IRIS is a NASA small explorer mission developed and operated by LMSAL with mission operations executed at NASA Ames Research Center and major contributions to downlink communications funded by ESA and the Norwegian Space Centre. We sincerely thank the Hinode team and the IRIS team for providing the coordinated observations used in this article.
This work was partially supported by JSPS KAKENHI Grant No. JP18H05234.
\end{acknowledgments}

%

\vspace{5mm}








\bibliography{sample631}{}

\begin{thebibliography}{}
\expandafter\ifx\csname natexlab\endcsname\relax\def\natexlab#1{#1}\fi

\bibitem[{{Abbasvand} {et~al.}(2020{\natexlab{a}}){Abbasvand}, {Sobotka},
  {Heinzel}, {{\v{S}}vanda}, {Jur{\v{c}}{\'a}k}, {del Moro}, \&
  {Berrilli}}]{Abbasvand2020b}
{Abbasvand}, V., {Sobotka}, M., {Heinzel}, P., {et~al.} 2020{\natexlab{a}},
  ApJ, 890, 22

\bibitem[{{Abbasvand} {et~al.}(2020{\natexlab{b}}){Abbasvand}, {Sobotka},
  {{\v{S}}vanda}, {Heinzel}, {Garc{\'\i}a-Rivas}, {Denker}, {Balthasar},
  {Verma}, {Kontogiannis}, {Koza}, {Korda}, \& {Kuckein}}]{Abbasvand2020a}
{Abbasvand}, V., {Sobotka}, M., {{\v{S}}vanda}, M., {et~al.}
  2020{\natexlab{b}}, A\&A, 642, A52

\bibitem[{{Barthol} {et~al.}(2011){Barthol}, {Gandorfer}, {Solanki},
  {Sch{\"u}ssler}, {Chares}, {Curdt}, {Deutsch}, {Feller}, {Germerott},
  {Grauf}, {Heerlein}, {Hirzberger}, {Kolleck}, {Meller}, {M{\"u}ller},
  {Riethm{\"u}ller}, {Tomasch}, {Kn{\"o}lker}, {Lites}, {Card}, {Elmore},
  {Fox}, {Lecinski}, {Nelson}, {Summers}, {Watt}, {Mart{\'\i}nez Pillet},
  {Bonet}, {Schmidt}, {Berkefeld}, {Title}, {Domingo}, {Gasent Blesa}, {Del
  Toro Iniesta}, {L{\'o}pez Jim{\'e}nez}, {{\'A}lvarez-Herrero},
  {Sabau-Graziati}, {Widani}, {Haberler}, {H{\"a}rtel}, {Kampf}, {Levin},
  {P{\'e}rez Grande}, {Sanz-Andr{\'e}s}, \& {Schmidt}}]{Barthol2011}
{Barthol}, P., {Gandorfer}, A., {Solanki}, S.~K., {et~al.} 2011, Sol. Phys.,
  268, 1

\bibitem[{{Bel} \& {Leroy}(1977)}]{Bel1977}
{Bel}, N., \& {Leroy}, B. 1977, A\&A, 55, 239

\bibitem[{{Bloomfield} {et~al.}(2007){Bloomfield}, {Lagg}, \&
  {Solanki}}]{Bloomfield2007}
{Bloomfield}, D., {Lagg}, A., \& {Solanki}, S. 2007, ApJ, 671, 1005

\bibitem[{{Cally} \& {Moradi}(2013)}]{Cally2013}
{Cally}, P., \& {Moradi}, H. 2013, MNRAS, 435, 2589

\bibitem[{{Carlsson} {et~al.}(2016){Carlsson}, {Hansteen}, {Gudiksen},
  {Leenaarts}, \& {De Pontieu}}]{Carlsson2016}
{Carlsson}, M., {Hansteen}, V.~H., {Gudiksen}, B.~V., {Leenaarts}, J., \& {De
  Pontieu}, B. 2016, A\&A, 585, A4

\bibitem[{{Carlsson} {et~al.}(2015){Carlsson}, {Leenaarts}, \& {De
  Pontieu}}]{Carlsson2015}
{Carlsson}, M., {Leenaarts}, J., \& {De Pontieu}, B. 2015, ApJL, 809, L30

\bibitem[{{Cavallini}(2006)}]{Cavallini2006}
{Cavallini}, F. 2006, Sol. Phys., 236, 415

\bibitem[{{Centeno} {et~al.}(2009){Centeno}, {Collados}, \& {Trujillo
  Bueno}}]{Centeno2009}
{Centeno}, R., {Collados}, M., \& {Trujillo Bueno}, J. 2009, ApJ, 692, 1211

\bibitem[{{Chae} {et~al.}(2013){Chae}, {Park}, \& {Ahn}}]{Chae2013}
{Chae}, J., {Park}, H.-M., \& {Ahn}, K. e.~a. 2013, Solar Physics, 288, 1

\bibitem[{{De Pontieu} {et~al.}(2014){De Pontieu}, {Rouppe van der Voort},
  {Pereira}, {Skogsrud}, {McIntosh}, {Carlsson}, \& {Hansteen}}]{DePontieu2014}
{De Pontieu}, B., {Rouppe van der Voort}, L., {Pereira}, T. M.~D., {et~al.}
  2014, in American Astronomical Society Meeting Abstracts, Vol. 224, American
  Astronomical Society Meeting Abstracts \#224, 313.02

\bibitem[{{Felipe} {et~al.}(2018){Felipe}, {Kuckein}, \& {Thaler}}]{Felipe2018}
{Felipe}, T., {Kuckein}, C., \& {Thaler}, I. 2018, A\&A, 617, A39

\bibitem[{{Felipe} \& {Sangeetha}(2020)}]{Felipe2020}
{Felipe}, T., \& {Sangeetha}, C.~R. 2020, A\&A, 640, A4

\bibitem[{{Graham} \& {Cauzzi}(2015)}]{Graham2015}
{Graham}, D.~R., \& {Cauzzi}, G. 2015, APJL, 807, L22

\bibitem[{{Gudiksen} {et~al.}(2011){Gudiksen}, {Carlsson}, {Hansteen}, {Hayek},
  {Leenaarts}, \& {Mart{\'\i}nez-Sykora}}]{Gudiksen2011}
{Gudiksen}, B.~V., {Carlsson}, M., {Hansteen}, V.~H., {et~al.} 2011, A\&AP,
  531, A154

\bibitem[{{Hasegawa}(2021)}]{Hasegawa2021}
{Hasegawa}, T. 2021, PhD thesis, The University of Tokyo

\bibitem[{{Heggland} {et~al.}(2011){Heggland}, {Hansteen}, {De Pontieu}, \&
  {Carlsson}}]{Heggland2011}
{Heggland}, L., {Hansteen}, V.~H., {De Pontieu}, B., \& {Carlsson}, M. 2011,
  ApJ, 743, 142

\bibitem[{{Ichimoto} {et~al.}(2008){Ichimoto}, {Lites}, {Elmore}, {Suematsu},
  {Tsuneta}, {Katsukawa}, {Shimizu}, {Shine}, {Tarbell}, {Title}, {Kiyohara},
  {Shinoda}, {Card}, {Lecinski}, {Streander}, {Nakagiri}, {Miyashita},
  {Noguchi}, {Hoffmann}, \& {Cruz}}]{Ichimoto2008}
{Ichimoto}, K., {Lites}, B., {Elmore}, D., {et~al.} 2008, Solar Physics, 249,
  233

\bibitem[{{Jefferies} {et~al.}(2006){Jefferies}, {McIntosh}, {Armstrong},
  {Bogdan}, {Cacciani}, \& {Fleck}}]{Jefferies2006}
{Jefferies}, S.~M., {McIntosh}, S.~W., {Armstrong}, J.~D., {et~al.} 2006, ApJL,
  648, L151

\bibitem[{{Jess}(2023)}]{Jess2023}
{Jess}, D.~B. 2023, Living Reviews in Solar Physics, 20, 1

\bibitem[{{Jess} {et~al.}(2013){Jess}, E., {van Doorsselaere}, {Keys}, \&
  {Mackay}}]{Jess2013}
{Jess}, D.~B., E., R.~V., {van Doorsselaere}, T., {Keys}, P.~H., \& {Mackay},
  D.~H. 2013, ApJ, 779, 168

\bibitem[{{Kanoh} {et~al.}(2016){Kanoh}, {Shimizu}, \& {Imada}}]{Kanoh2016}
{Kanoh}, R., {Shimizu}, T., \& {Imada}, S. 2016, ApJ, 831, 24

\bibitem[{{Katsukawa} {et~al.}(2020){Katsukawa}, {del Toro Iniesta}, {Solanki},
  {Kubo}, {Hara}, {Shimizu}, {Oba}, {Kawabata}, {Tsuzuki}, {Uraguchi},
  {Nodomi}, {Shinoda}, {Tamura}, {Suematsu}, {Ishikawa}, {Kano}, {Matsumoto},
  {Ichimoto}, {Nagata}, {Quintero Noda}, {Anan}, {Orozco Su{\'a}rez}, {Balaguer
  Jim{\'e}nez}, {L{\'o}pez Jim{\'e}nez}, {Cobos Carrascosa}, {Feller},
  {Riethmueller}, {Gandorfer}, \& {Lagg}}]{Katsukawa2020}
{Katsukawa}, Y., {del Toro Iniesta}, J.~C., {Solanki}, S.~K., {et~al.} 2020, in
  Society of Photo-Optical Instrumentation Engineers (SPIE) Conference Series,
  Vol. 11447, Society of Photo-Optical Instrumentation Engineers (SPIE)
  Conference Series, 114470Y

\bibitem[{{Kosugi} {et~al.}(2007){Kosugi}, {Matsuzaki}, {Sakao}, {Shimizu},
  {Sone}, {Tachikawa}, {Hashimoto}, {Minesugi}, {Ohnishi}, {Yamada}, {Tsuneta},
  {Hara}, {Ichimoto}, {Suematsu}, {Shimojo}, {Watanabe}, {Shimada}, {Davis},
  {Hill}, {Owens}, {Title}, {Culhane}, {Harra}, {Doschek}, \&
  {Golub}}]{Kosugi2007}
{Kosugi}, T., {Matsuzaki}, K., {Sakao}, T., {et~al.} 2007, Sol. Phys., 243, 3

\bibitem[{{Leenaarts} {et~al.}(2013{\natexlab{a}}){Leenaarts}, {Pereira},
  {Carlsson}, {Uitenbroek}, \& {De Pontieu}}]{Leenaarts2013a}
{Leenaarts}, J., {Pereira}, T.~M.~D., {Carlsson}, M., {Uitenbroek}, H., \& {De
  Pontieu}, B. 2013{\natexlab{a}}, ApJ, 772, 89

\bibitem[{{Leenaarts} {et~al.}(2013{\natexlab{b}}){Leenaarts}, {Pereira},
  {Carlsson}, {Uitenbroek}, \& {De Pontieu}}]{Leenaarts2013b}
---. 2013{\natexlab{b}}, ApJ, 772, 90

\bibitem[{{Lites} \& {Ichimoto}(2013)}]{Lites2013b}
{Lites}, B.~W., \& {Ichimoto}, K. 2013, Solar Physics, 283, 601

\bibitem[{{Lites} {et~al.}(2013){Lites}, {Akin}, {Card}, {Cruz}, {Duncan},
  {Edwards}, {Elmore}, {Hoffmann}, {Katsukawa}, {Katz}, {Kubo}, {Ichimoto},
  {Shimizu}, {Shine}, {Streander}, {Suematsu}, {Tarbell}, {Title}, \&
  {Tsuneta}}]{Lites2013}
{Lites}, B.~W., {Akin}, D.~L., {Card}, G., {et~al.} 2013, Sol. Phys., 283, 579

\bibitem[{{L{\"o}hner-B{\"o}ttcher} \& {Bello
  Gonz{\'a}lez}(2015)}]{Lohner-Bottcher2015}
{L{\"o}hner-B{\"o}ttcher}, J., \& {Bello Gonz{\'a}lez}, N. 2015, A\&A, 580, A53

\bibitem[{{McIntosh} \& {Jefferies}(2006)}]{McIntosh2006}
{McIntosh}, S.~W., \& {Jefferies}, S.~M. 2006, ApJL, 647, L77

\bibitem[{{Morrill} {et~al.}(2001){Morrill}, {Dere}, \&
  {Korendyke}}]{Morrill2001}
{Morrill}, J.~S., {Dere}, K.~P., \& {Korendyke}, C.~M. 2001, ApJ, 557, 854

\bibitem[{{Oba} {et~al.}(2017){Oba}, {Iida}, \& {Shimizu}}]{Oba2017}
{Oba}, T., {Iida}, Y., \& {Shimizu}, T. 2017, ApJ, 836, 40

\bibitem[{{Parker}(1972)}]{Parker1972}
{Parker}, E.~N. 1972, ApJ, 174, 499

\bibitem[{{Parker}(1988)}]{Parker1988}
---. 1988, ApJ, 330, 474

\bibitem[{{Pereira} {et~al.}(2013){Pereira}, {Leenaarts}, {De Pontieu},
  {Carlsson}, \& {Uitenbroek}}]{Pereira2013}
{Pereira}, T.~M.~D., {Leenaarts}, J., {De Pontieu}, B., {Carlsson}, M., \&
  {Uitenbroek}, H. 2013, ApJ, 778, 143

\bibitem[{{Pontin} \& {Hornig}(2020)}]{Pontin2020}
{Pontin}, D., \& {Hornig}, G. 2020, Living Reviews in Solar Physics, 17, 5

\bibitem[{{Priest}(2014)}]{Priest2014}
{Priest}, E. 2014, {Magnetohydrodynamics of the Sun} (Cambridge University
  Press), doi:https://doi.org/10.1017/CBO9781139020732

\bibitem[{{Quintero Noda} {et~al.}(2017){Quintero Noda}, {Shimizu}, \&
  {Katsukawa}}]{QuinteroNoda2017}
{Quintero Noda}, C., {Shimizu}, T., \& {Katsukawa}, Y. e.~a. 2017, MNRAS, 464,
  4534

\bibitem[{{Schmit} {et~al.}(2015){Schmit}, {Bryans}, {De Pontieu}, {McIntosh},
  {Leenaarts}, \& {Carlsson}}]{Schmit2015}
{Schmit}, D., {Bryans}, P., {De Pontieu}, B., {et~al.} 2015, ApJ, 811, 127

\bibitem[{{Schmitz} \& {Fleck}(1998)}]{Schmitz1998}
{Schmitz}, F., \& {Fleck}, B. 1998, A\&A, 337, 487

\bibitem[{{Schunker} \& {Cally}(2006)}]{Schunker2006}
{Schunker}, H., \& {Cally}, P. 2006, MNRAS, 372, 551

\bibitem[{{Shimizu} {et~al.}(2008){Shimizu}, {Nagata}, {Tsuneta}, {Tarbell},
  {Edwards}, {Shine}, {Hoffmann}, {Thomas}, {Sour}, {Rehse}, {Ito},
  {Kashiwagi}, {Tabata}, {Kodeki}, {Nagase}, {Matsuzaki}, {Kobayashi},
  {Ichimoto}, \& {Suematsu}}]{Shimizu2008}
{Shimizu}, T., {Nagata}, S., {Tsuneta}, S., {et~al.} 2008, Solar Physics, 249,
  221

\bibitem[{{Sobotka} {et~al.}(2016){Sobotka}, {Heinzel}, {{\v{S}}vanda},
  {Jur{\v{c}}{\'a}k}, {del Moro}, \& {Berrilli}}]{Sobotka2016}
{Sobotka}, M., {Heinzel}, P., {{\v{S}}vanda}, M., {et~al.} 2016, ApJ, 826, 49

\bibitem[{{Suematsu} {et~al.}(2008){Suematsu}, {Tsuneta}, {Ichimoto},
  {Shimizu}, {Otsubo}, {Katsukawa}, {Nakagiri}, {Noguchi}, {Tamura}, {Kato},
  {Hara}, \& {Kubo}}]{Suematsu2008}
{Suematsu}, Y., {Tsuneta}, S., {Ichimoto}, K., {et~al.} 2008, Solar Physics,
  249, 197

\bibitem[{{Title} {et~al.}(1989){Title}, {Tarbell}, {Topka}, {Ferguson},
  {Shine}, \& {SOUP Team}}]{Title1989}
{Title}, A.~M., {Tarbell}, T.~D., {Topka}, K.~P., {et~al.} 1989, \apj, 336, 475

\bibitem[{{Tsuneta} {et~al.}(2008){Tsuneta}, {Ichimoto}, {Katsukawa}, {Nagata},
  {Otsubo}, {Shimizu}, {Suematsu}, {Nakagiri}, {Noguchi}, {Tarbell}, {Title},
  {Shine}, {Rosenberg}, {Hoffmann}, {Jurcevich}, {Kushner}, {Levay}, {Lites},
  {Elmore}, {Matsushita}, {Kawaguchi}, {Saito}, {Mikami}, {Hill}, \&
  {Owens}}]{Tsuneta2008}
{Tsuneta}, S., {Ichimoto}, K., {Katsukawa}, Y., {et~al.} 2008, Sol. Phys., 249,
  167

\bibitem[{{Uitenbroek}(2001)}]{Uitenbroek2001}
{Uitenbroek}, H. 2001, ApJ, 557, 389

\bibitem[{{Vernazza} {et~al.}(1981){Vernazza}, {Avrett}, \&
  {Loeser}}]{Vernazza1981}
{Vernazza}, J.~E., {Avrett}, E.~H., \& {Loeser}, R. 1981, ApJS, 45, 635

\bibitem[{{von der L{\"u}he}(1998)}]{vonderLuhe1998}
{von der L{\"u}he}, O. 1998, New Astron. Rev., 42, 493

\bibitem[{{Withbroe} \& {Noyes}(1977)}]{Withbroe1977}
{Withbroe}, G.~L., \& {Noyes}, R.~W. 1977, ARA\&A, 15, 363

\end{thebibliography}
\bibliographystyle{aasjournal}



\end{document}